 \documentclass[final,3p,times,twocolumn]{elsarticle}

\usepackage{amsmath, amsfonts, latexsym, amssymb, graphicx, graphics, epsfig, subfigure, color, makeidx}
\usepackage{xcolor, epstopdf}
\usepackage{microtype}
\usepackage{ulem}
\usepackage{siunitx}

\usepackage{gensymb}
\usepackage{float} 
\DeclareSIUnit[number-unit-product = {}]{\inchQ}{\textquotedbl}
\DeclareSIUnit[number-unit-product = {\thinspace}]{\inch}{in}
\usepackage[colorlinks]{hyperref}
\usepackage{caption}
\usepackage{appendix}
\usepackage{dcolumn}
\usepackage{bm}
\usepackage{float}
\usepackage{multirow}
\usepackage{longtable}
\usepackage{dcolumn}
\newcolumntype{d}{D{.}{.}{-1}}
\usepackage{setspace}
\usepackage{threeparttable}
\usepackage{tabularx}
\usepackage[symbol*]{footmisc}
\DefineFNsymbolsTM{myfnsymbols}{
  \textasteriskcentered *
  \textdagger    \dagger
  \textdaggerdbl \ddagger
  \textsection   \mathsection
  \textbardbl    \|%
  \textparagraph \mathparagraph
}%
\setfnsymbol{myfnsymbols}

\usepackage{gensymb}

\biboptions{sort,compress}

\begin{document}

\begin{frontmatter}



\title{\boldmath Different effects of the Lorentz and Gaussian bump functions on the formation of primordial black holes and secondary gravitational waves}


\author[loc1]{Wei Yang}
\ead{yw1214@nuaa.edu.cn}

\author[loc2]{Yu-Xuan Kang}

\author[loc3]{Arshad Ali}

\author[loc1]{Tao-Tao Sui}

\author[loc1]{Chen-Hao Wu}

\author[loc1,loc4]{Ya-Peng Hu\corref{cor1}}
\ead{huyp@nuaa.edu.cn}
\cortext[cor1]{Corresponding author}

\address[loc1]{College of Physics, Nanjing University of Aeronautics and Astronautics, Nanjing, 211106, China}
\address[loc2]{School of Physics Sciences, University of Science and Technology of China, Hefei 230026, China}
\address[loc3]{Institute for Advanced Study \& School of Physical Science and Technology, Soochow University, Suzhou 215006, P.R. China}
\address[loc4]{Key Laboratory of Aerospace Information Materials and Physics (NUAA), MIIT, Nanjing 211106, China}







\begin{abstract}{Scalar perturbations in the inflation can be amplified when the base inflation potential $V_b(\phi)$ incorporates a local bump $f(\phi)$ such as $V(\phi)=V_b(\phi)(1+f(\phi))$. This modification will lead to a peak in the curvature power spectrum, increasing a significant abundance of primordial black holes (PBHs). However, since there is no underlying physical reason for the choice of $f(\phi)$, it is essential to investigate the effects of various bump functions on PBH generation. In this paper, we choose the well-known Starobinsky potential as the base inflation potential to compare the effects produced by different bumps, specifically focusing on the Lorentz and Gaussian bumps which are widely used. To clearly illustrate the differences between these two bumps, we keep parameters in bump functions the same. We find an interesting and novel result that the Lorentz cases manifest a stronger ability to enhance the power spectrum and produce more abundance of PBHs than Gaussian cases.
Moreover, we also investigate the different effects of bump functions on the scalar-induced gravitational waves (SIGWs). The results indicate that the Lorentz bump generates SIGWs with a higher energy density, which can be potentially detected in the future. Our study gives valuable insights into the choice and constraints on the bump functions, and the different effects may distinguish the two bump cases for practical purposes in future experiments.}

\end{abstract}


\end{frontmatter}

\section{Introduction}\label{introduction}
Since gravitational waves (GWs) were first detected in September 2015, the Laser Interferometer Gravitational Wave Observatory (LIGO)/Virgo Collaboration has identified gravitational wave signals from ten binary black hole (BH) mergers and one binary neutron star merger \cite{LIGOScientific:2016aoc, LIGOScientific:2018mvr}. It is worth noting that PBH has also attracted extensive attention due to its possibility to explain this merger event, and has become one of the important candidate for explaining the origin of related GW events at present\cite{Bird:2016dcv, Sasaki:2016jop}.

PBHs have garnered considerable attention as viable candidates for cold dark matter (DM)~\cite{Kovetz:2017rvv, Carr:2016drx, Carr:2020gox}. Moreover, several astrophysical observations provide credible evidence that DM is a non-negligible element of our universe \cite{Chapline:1975ojl, Persic:1995ru, Bertone:2016nfn}. Therefore, the abundance of primordial black holes as dark matter is a critical issue investigated in cosmological and astronomical observations~\cite{Ali-Haimoud:2016mbv, Poulin:2017bwe, Carr:2009jm, Niikura:2017zjd, Griest:2013esa, EROS-2:2006ryy, Vaskonen:2019jpv}.  However, the abundances of PBHs within the mass ranges of $10^{-17}-10^{-15}M_{\odot}$ and $10^{-14}-10^{-12}M_{\odot}$ are largely unconstrained by observations and specific assumptions, which may make up almost all of the DM \cite{Green:2020jor}. 

Several mechanisms have been proposed for the formation of PBHs. The most popular one is the gravitational collapse of the overdense region inside a Hubble horizon, where the density exceeds the critical value \cite{Hawking:1971ei}. More recently, some people also investigated from the viewpoint of compaction function \cite{Musco:2018rwt, Germani:2018jgr, Germani:2019zez, Escriva:2019phb}. Many situations can form the overdense region, such as enhancement of the scalar perturbation \cite{Ivanov:1994pa, Garcia-Bellido:2017mdw, Sasaki:2018dmp, Yi:2020cut, Pi:2021dft, Wu:2021zta, Kawai:2021edk, Kawai:2022emp, Germani:2017bcs, Motohashi:2017kbs, Atal:2019erb, Ballesteros:2017fsr}, accumulation of topological defect and postponed false vacuum decay during the first-order transition \cite{Deng:2016vzb, Eroshenko:2021sez, Lewicki:2023ioy, Liu:2021svg}. Besides the formation of PBHs, the scalar-induced gravitational waves (SIGWs) are also produced during perturbations \cite{Wang:2016ana, Cai:2018dig, Inomata:2023zup, Kohri:2020qqd, Basilakos:2023xof}, which are expected to test by the North American Nanohertz Observatory for Gravitational Wave (NANOGrav) \cite{NANOGrav:2023gor}, Pulsar Timing Arrays (PTA) \cite{EPTA:2023fyk, Reardon:2023gzh} and space-based GW detectors, such as the Laser Interferometer Space Antenna (LISA) \cite{LISA:2017pwj}, Taiji \cite{Hu:2017mde}, TianQin \cite{TianQin:2015yph}, and Deci-hertz Interferometer Gravitational-Wave Observatory (DECIGO) \cite{Kawamura:2006up} in the future. To achieve a sufficient abundance of PBHs as DM and detectable SIGWs, a large density perturbation at a small scale is typically required. 

Recently, with the aim of enhancing
the scalar perturbation to generate PBHs, researchers found that one could also achieve this enhancement in the single-field inflation process by involving a local bump on the base inflation potential function~\cite{Wang:2021kbh, Mishra:2019pzq, Motohashi:2017kbs, Atal:2019erb}. That is, the inflation potential $V(\phi)$ is small modified on the base potential $V_\text{b}(\phi)$ like $V(\phi)=V_\text{b}(\phi)(1+f(\phi))$, where $f(\phi)$ represents a local correction bump function. The local bump usually produces a notable peak in the curvature power spectrum, and hence can increase the probability of PBH formation. Note that, various local bump functions have been chosen such as the Gaussian bump, hyperbolic bump, and Lorentz bump, while there is still a lack of underlying reasons for the choice of bump functions. Therefore, it will be illuminated that one investigates different properties of bump functions adopting the same base inflation potential. 

In our paper, we choose the Starobinsky potential as the same base inflation potential, since the Starobinsky model is a well-known inflation model and has been verified to be compatible well with cosmic microwave background (CMB) observations until now \cite{Starobinsky:1980te, Mishra:2018dtg}. Furthermore, for simplicity, we mainly focus on the comparison between the Lorentz bump and Gaussian bump, since both of them are widely used in the modified potential $V(\phi)$~\cite{Mishra:2019pzq, Wang:2021kbh, Motohashi:2017kbs}. In addition, in order to clearly illustrate the difference between these two bumps, we keep parameters $b, \phi_0, c$ in bump functions the same. In this case, the shapes of Lorentz and Gaussian functions are similar, i.e., the amplitude and peak position are the same excluding the width of these two functions are different. However, the Lorentz bump expresses a stronger ability to enhance the power spectrum, and produces more abundance of PBHs and a larger energy density of SIGWs. These differences in our results shed insights into the choice and understanding of these two bump functions.

This paper is organized as follows. In section \ref{sec:2}, we review the Starobinsky inflation model and incorporate Lorentz and Gaussian bump functions. In section \ref{sec:3}, we express the different effects of Lorentz and Gaussian cases on the power spectrum. In section \ref{sec:4}, we obtain the abundance of PBHs in the two bump models. In section \ref{sec:5}, we investigate the energy density spectrum of SIGWs. We draw our conclusion and discussion in section \ref{sec:6}.

\section{Starobinsky inflation potential with different bump forms}\label{sec:2}
In this paper, we study the single field inflation with the corresponding action
\begin{gather}
S=\int d^4x\sqrt{-g}\left[\frac{1}{2}R-\frac{1}{2}g_{\mu\nu}\nabla^\mu\phi\nabla^\nu\phi-V(\phi)\right].
\end{gather}

In the context of a spatially flat homogeneous and isotropic universe with Flat-Lemaitre-Robertson-Walker (FLRW) metric $ds^2=-dt^2+a(t)\delta_\text{ij}dx^\text{i}dx^\text{j}$, the Friedmann equations and dynamical equation of the inflation field $\phi$ are~\cite{Garcia-Bellido:2017mdw}
\begin{align}
3H^2=\frac{1}{2}\dot{\phi}^2+V(\phi),\nonumber\\
\dot{H}=-\frac{1}{2}\dot{\phi}^2,\nonumber\\
\ddot{\phi}+3H\dot{\phi}+V_\phi(\phi)=0,\label{Friedman}
\end{align}
where $a(t)$ is the scale factor, and $H=\dot{a}/a$ represents the Hubble parameter. The `dot' here is the derivative of $t$ in our paper. The $V_\phi(\phi)$ is the derivative of the potential function $V(\phi)$ with respect to $\phi$.

For inflation models, two slow-roll parameters can be defined as
\begin{gather}\label{srh}
\epsilon_\text{H}=-\frac{\dot{H}}{H^2},~~~~~~ \eta_\text{H}=-\frac{\ddot{\phi}}{H\dot{\phi}}. 
\end{gather}
In the slow-roll inflation process, the two parameters satisfy conditions $\epsilon_\text{H}\ll1$ and $|\eta_\text{H}|\ll1$, and the end of inflation condition is $\epsilon_\text{H}=1$. Moreover, the total e-folding number $N$ can be written according to the slow-roll parameter $\epsilon_\text{H}$ as
\begin{align}
\label{N}
N=\int_{\phi_\text{in}}^{\phi_\text{e}}\frac{1}{\sqrt{2\epsilon_\text{H}}}d\phi,
\end{align}
where $\phi_\text{in}$ represents the initial value of the scalar field in inflation, and $\phi_\text{e}$ is the final value at the end of inflation. 

During the inflation epoch, the slow-roll conditions are equivalent to $\frac{1}{2}\dot{\phi}^2\ll V(\phi)$ and $|\ddot{\phi}| \ll 3H|\dot{\phi}|$. Under these conditions, the Friedmann equations and dynamical equation are approximately written as \cite{Yi:2020cut}
\begin{align}
\label{slow approximation}
3H^2\simeq V(\phi),~~~~3H\dot{\phi}\simeq-V_\phi(\phi).
\end{align}
With these two equations, one can obtain the well-known slow-roll approximation formula of the power spectrum for scalar perturbation as \cite{Garcia-Bellido:2017mdw}
\begin{gather}
\label{Psr}
P_{\zeta}=\frac{H^2}{8\pi^2\epsilon_\text{H}}.
\end{gather}
which implies that one can enhance the power spectrum by reducing the slow roll parameter $\epsilon_\text{H}$. Note that, in the slow-roll approximation, the slow roll parameter $\epsilon_\text{H}$ satisfies $\epsilon_H\simeq \frac{1}{2}\left(\frac{V_{\phi}(\phi)}{V(\phi)}\right)^2$. If $V_{\phi}(\phi)$ is close to 0, the $V(\phi)$ and $H$ are nearly a constant from Eq. \eqref{slow approximation}. Then, the slow roll parameter $\epsilon_\text{H}$ will decrease to nearly zero by several orders. Naturally, the power spectrum indeed can be enhanced from Eq. \eqref{Psr}.

In order to make the $V_{\phi}(\phi)$  close to 0, an approach has been proposed that one can add a Local tiny bump in the base inflation potential $V_\text{b}(\phi)$ such as\footnote{The potential $V(\phi)$ is an effective model from the phenomenology \cite{Mishra:2019pzq}, which may be considered as a small local radiative correction to the base potential.}\cite{Mishra:2019pzq}
\begin{equation}\label{V}
V(\phi)=V_\text{b}(\phi)(1+f(\phi)).
\end{equation}
where $f(\phi)$ appears as a tiny local bump feature and its value is much smaller than 1. In addition, the tiny bump function is often chosen as the Gaussian, hyperbolic, and Lorentz bumps. Since these functions have the maximum value at a point $\phi_0$, and hence can make $V_{\phi}(\phi)$ close to 0 around the point $\phi_0$.

Obviously, $V_{\phi}(\phi_0)$ is related to the choice of base inflation potential $V_\text{b}(\phi)$. In our paper, we choose the Starobinsky inflation potential as the base potential form written as
\begin{equation}\label{Vb}
V_\text{b}(\phi)=\frac{3}{4}m^2\left(1-e^{-\sqrt{\frac{2}{3}}\phi}\right)^2,
\end{equation}
since it is a well-known inflation model and has been verified to be compatible well with CMB observations until now~\cite{Planck:2018jri}. In addition, considering the constraint on the total number of e-foldings $N$ from CMB observations, one often assumes $N=60$. Therefore, we also adopt the same assumption for the base Starobinsky inflation potential with $N=60$, where one has chosen the mass parameter $m=1.13\times10^{-5}$, the initial and end values of the inflation field are $\phi_\text{in}=5.42$ and $\phi_\text{e}= 0.61394$, respectively~\cite{Mishra:2018dtg}.

Note that, after choosing the mass parameter, initial and end values of the inflation field, we can obtain the power spectrum $A_\textbf{s}\simeq 2.1\times 10^{-9}$ at the pivot scale $k_*=0.05\,\text{Mpc}^{-1}$, which satisfies the constraint of Planck 2018. Moreover, we obtain the scalar spectral index $n_{\text{s}}\simeq0.967$ and the tensor-to-scalar ratio $r\simeq0.003$ at the pivot scale $k_*=0.05\,\text{Mpc}^{-1}$ by using the equations $n_{\text{s}}=1+\frac{d\ln{P_\zeta}}{d\ln{k}}$ and $r=\frac{P_\text{T}}{P_{\mathcal{\zeta}}}$, which are consistent with the observational constraints 
$n_{\text{s}}=0.9649\pm0.0042 (68\% CL)$ \cite{Planck:2018jri} and $r_{0.05}<0.036$ at $95\%CL$ \cite{BICEP:2021xfz}. Here $P_\text{T}=2H^2/\pi^2$ is the power spectrum of tensor perturbation under the slow roll approximation, see Ref. \cite{DeFelice:2011zh} for details.

In this paper, we focus on the Lorentz function and Gaussian function as the local tiny bump added into the base Starobinsky potential, since these two bump functions are widely used in the modified potential $V(\phi)$ and may have some underlying physical meaning from the viewpoint of statistics. Then we mainly investigate the different effects of these two bump functions. 

We use the subscript L for the Lorentz function denoted by
\begin{equation}\label{Lorentz bump}
f_{\text{L}}(\phi)=\frac{b}{1+\left(\frac{\phi-\phi_\text{0}}{c_1}\right)^2},
\end{equation}
and adopt the subscript G for the Gaussian function written as
\begin{equation}\label{Gaussian bump}
f_\text{G}(\phi)=be^{-\frac{\left(\phi-\phi_\text{0}\right)^2}{2{c_2}^2}}.
\end{equation}
Note that, the parameters $b$ and $\phi_0$ control the amplitude and position in both bump functions respectively, thus we have kept these two parameters $b$ and $\phi_0$ same to illustrate the difference between these two bumps clearly. In addition, the parameters $c_1$ and $c_2$ are related to the width of bump functions. For simplicity, we also set $c_1=c_2$ and use a single parameter $c$ to represent the width in the following analysis of functions.

\begin{figure}[H]
\centering
\includegraphics[width=0.45\textwidth]{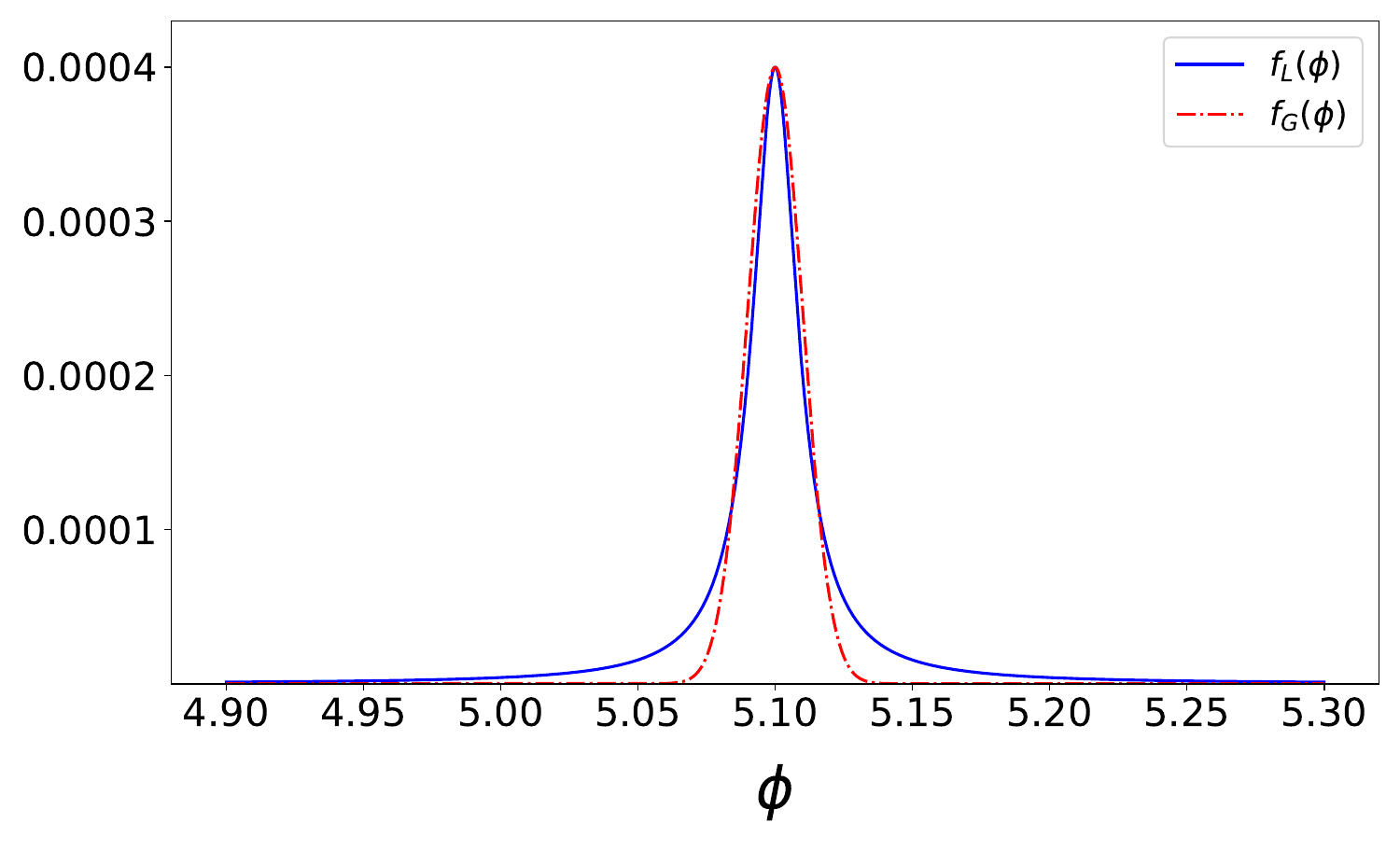}
\caption{Gaussian function and Lorentz function. We represent the Gaussian function and Lorentz function with red dotted lines and blue solid lines, respectively, where the parameter selection is $b=4\times10^{-4},c=0.00999227,\phi_0=5.1$.}\label{LG}
\end{figure}

\begin{figure}[H]
\centering
\includegraphics[width=0.45\textwidth]{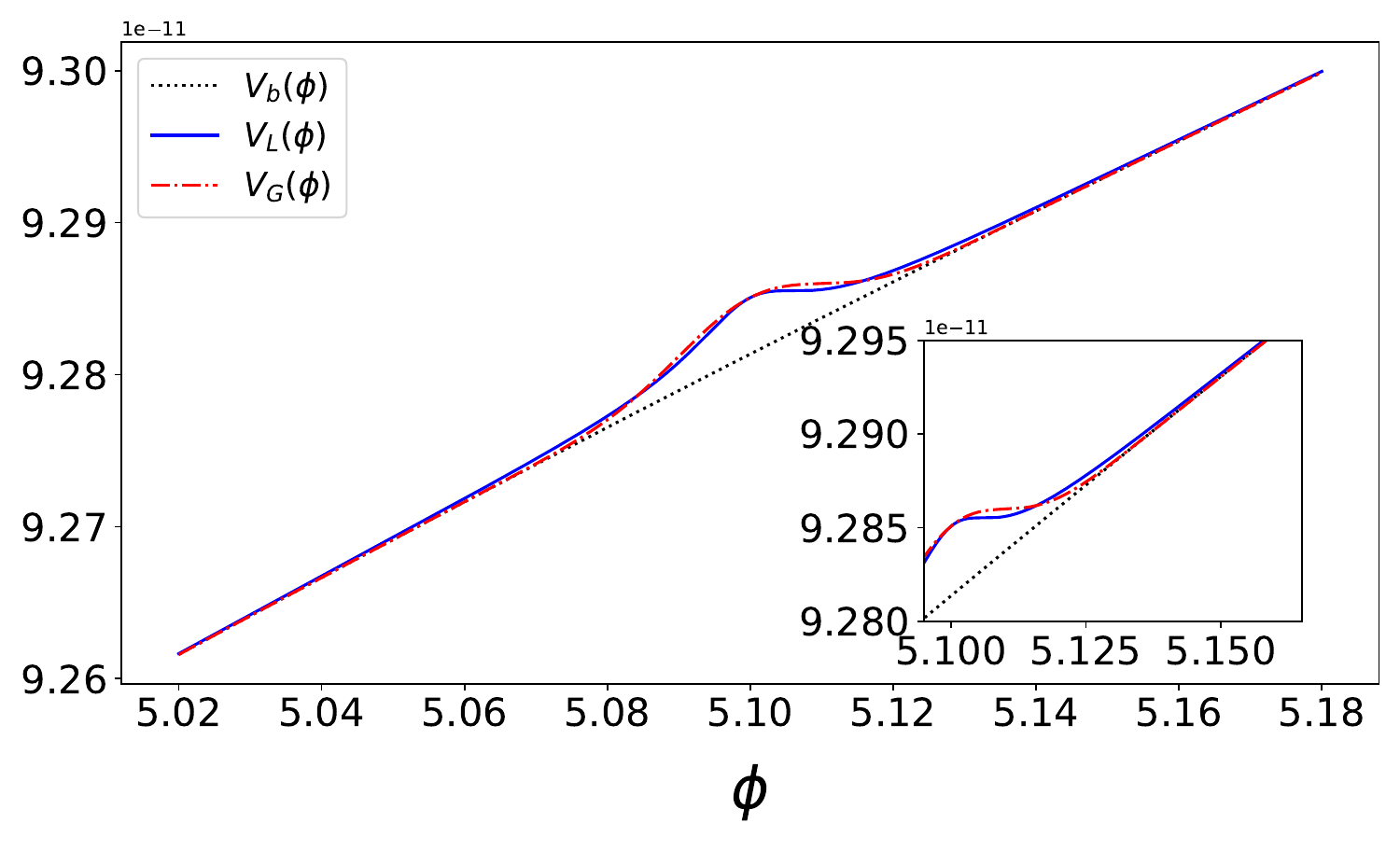}
\caption{The evolution of the potential functions with base potential (black dot), Gaussian bump (red dotted lines), and  Lorentz bump (blue solid lines), and the parameter selection is $b=4\times10^{-4},c=0.00999227,\phi_0=5.1$.}\label{V-bLG}
\end{figure}

To make the value of the bump function $f(\phi)$ much less than 1 with $f(\phi)\ll1$, we have chosen a set of values that satisfy this condition $b=4\times10^{-4}$, $c=0.00999227$ and $\phi_0=5.1$. Under this set of parameters, we have explicitly shown the Lorentz and Gaussian bump functions in Fig. \ref{LG}. We can easily find that the Lorentz bump function is slightly smoother than the Gaussian case due to the $f_{\text{G}}(\phi)$ decaying exponentially fast, while the $f_{\text{L}}(\phi)$ has a much slower power-law decay. This 'smoother' feature allows for a wider and flatter plateau in the potential, as shown in Fig. \ref{V-bLG}. This flat region leads to a much larger period of ultra-slow roll that ultimately leads to a more significant enhancement of the primordial perturbations, as we will demonstrate in the subsequent content.

\begin{figure}[H]
\centering
\subfigure{\includegraphics[width=0.45\textwidth]{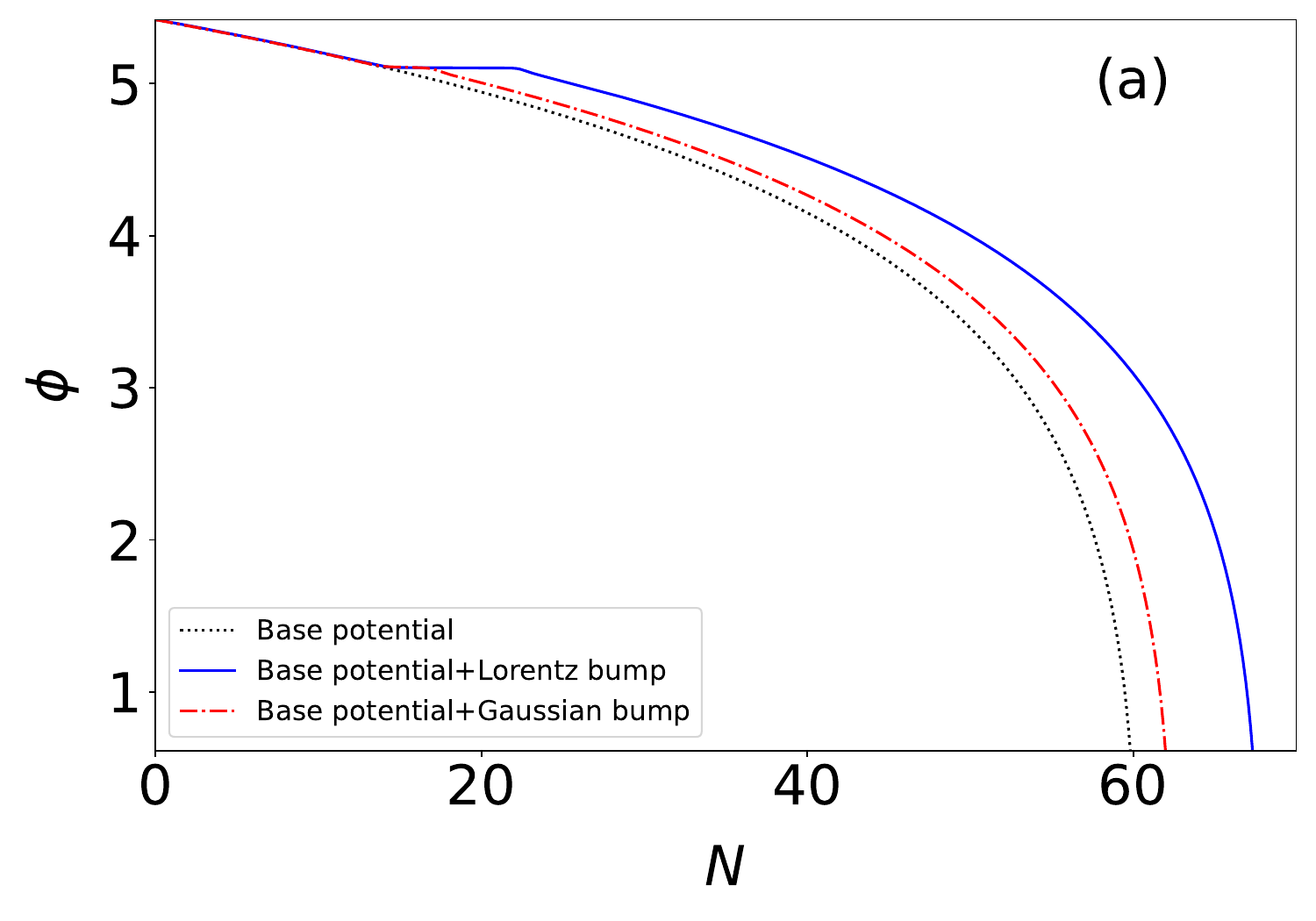}}\hfill
\subfigure{\includegraphics[width=0.45\textwidth]{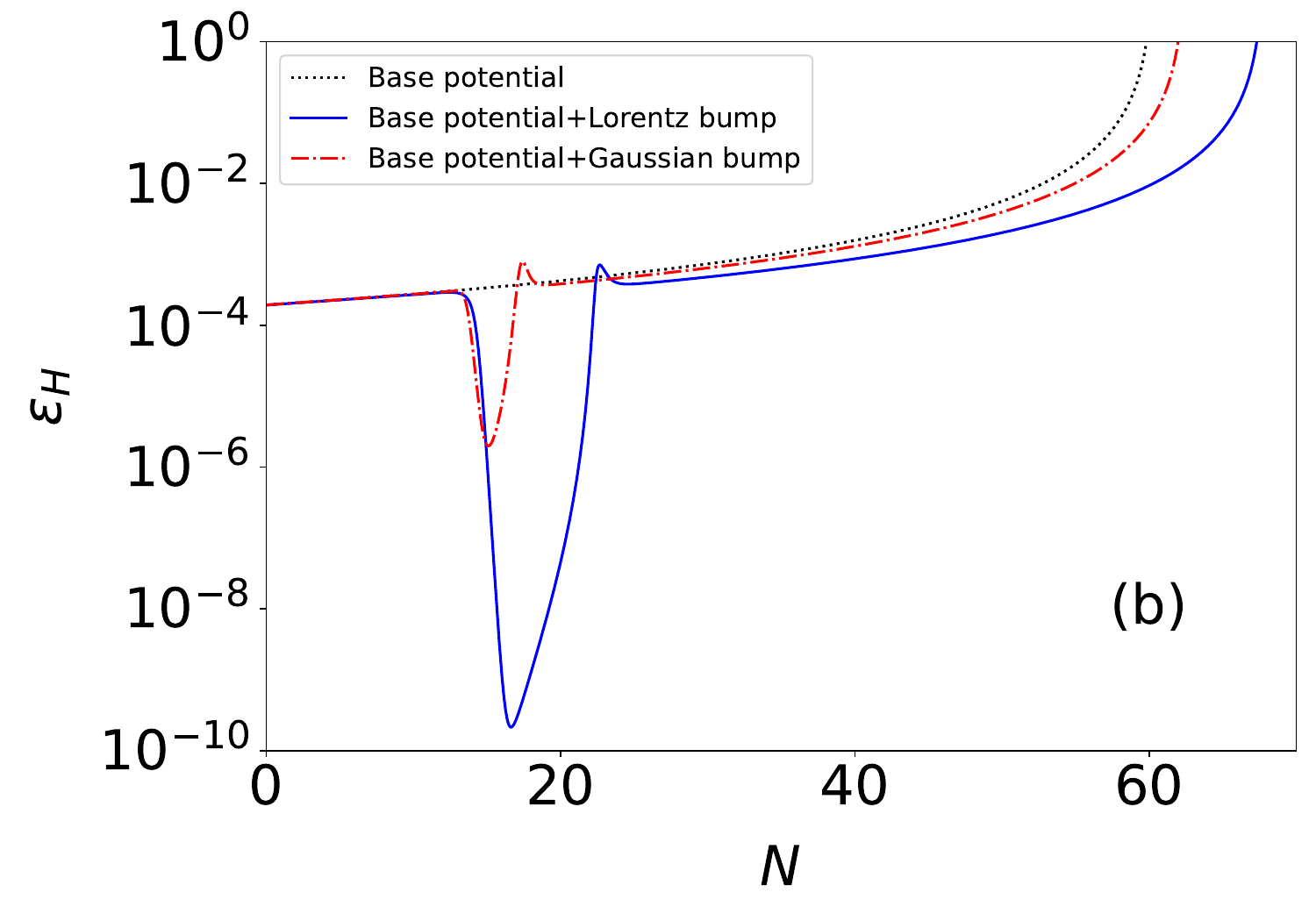}}
\caption{(a) shows the evolutions of e-folds number $N$. (b) shows the evolutions of the slow-roll parameter $\epsilon_\text{H}$, where the parameters of bump are $b=4\times10^{-4},c=0.00999227,\phi_0=5.1$.}\label{Nsrp}
\end{figure}

Now, we would like to extract the different effects of these two bump functions on the total e-folding numbers $N$. For convenience, we fix the initial value of the inflation field $\phi_\text{in}$ as $\phi_{in}=5.42$ even after adding the bump functions, which means that we assume the same initial value of the inflation field triggers the inflation process. After taking this assumption and $\phi_\text{e}$ satisfied condition $\epsilon_\text{H}=1$ into account, we indeed can illustrate the variation of total e-folding numbers in different cases, which has been shown in Fig. \ref{Nsrp}(a). Moreover, we also obtain the relationships between the slow-roll parameter $\epsilon_\text{H}$ and $N$ as shown in Fig. \ref{Nsrp}(b). 

From Fig. \ref{Nsrp}, we find that the Lorentz bump falls more slowly and is longer in the region near $\phi_0=5.1$, and hence the Lorentz bump provides larger e-folding numbers $N\sim67$ with $\phi_{\text{e}}=0.61435$, while $N\sim62$ with $\phi_{\text{e}}=0.61433$ for the Gaussian bump case. This difference arises because the “fatter tails” of Lorentz function cause a more profound and extended period of slow-roll, forcing the inflation to spend more e-folds traversing the region around $\phi_0 = 5.1$. Both scenarios remain consistent with observational constraints ($N\leq70$) \cite{Mishra:2019pzq, Yi:2020cut}. Moreover, we observe that the slow-roll parameter $\epsilon_\text{H}$ indeed decreases by several orders to nearly zero for both Gaussian and Lorentz bump functions, which will lead to the enhancement of the power spectrum from Eq. \eqref{Psr}. In the subsequent section, we will clearly illustrate these findings by numerically investigating the power spectrum. 

\section{The power spectrum of the Lorentz and Gaussian bumps models}\label{sec:3}

In this section, we mainly focus on comparing the behaviors of the enhanced power spectrum at small scales through differences exhibited from Lorentz and Gaussian cases. To achieve this, we employ the numerical method of solving the Mukhanov-Sasaki equation \cite{Mukhanov:1985rz, Sasaki:1986hm}
\begin{gather}\label{MS}
  \frac{d^2v_\text{k}}{d\eta^2}+\left(k^2-\frac{1}{z}\frac{d^2z}{d\eta^2}\right)v_\text{k}=0,
\end{gather}
where the conformal time $\eta$ is satisfied $d\eta=dt/a$, and the quantum canonical field is defined as $v_\text{k}=z\zeta_\text{k}$ with $z=a\dot{\phi}/H$ and comoving curvature perturbation $\zeta_\text{k}$. Through numerically obtaining the $v_\text{k}$, we can calculate the scalar power spectrum by
\begin{equation}\label{P}
P_{\zeta}=\frac{k^3}{2\pi^2}|\zeta_\text{k}|^2.
\end{equation}

\begin{figure}[H]
\centering
\includegraphics[width=0.45\textwidth]{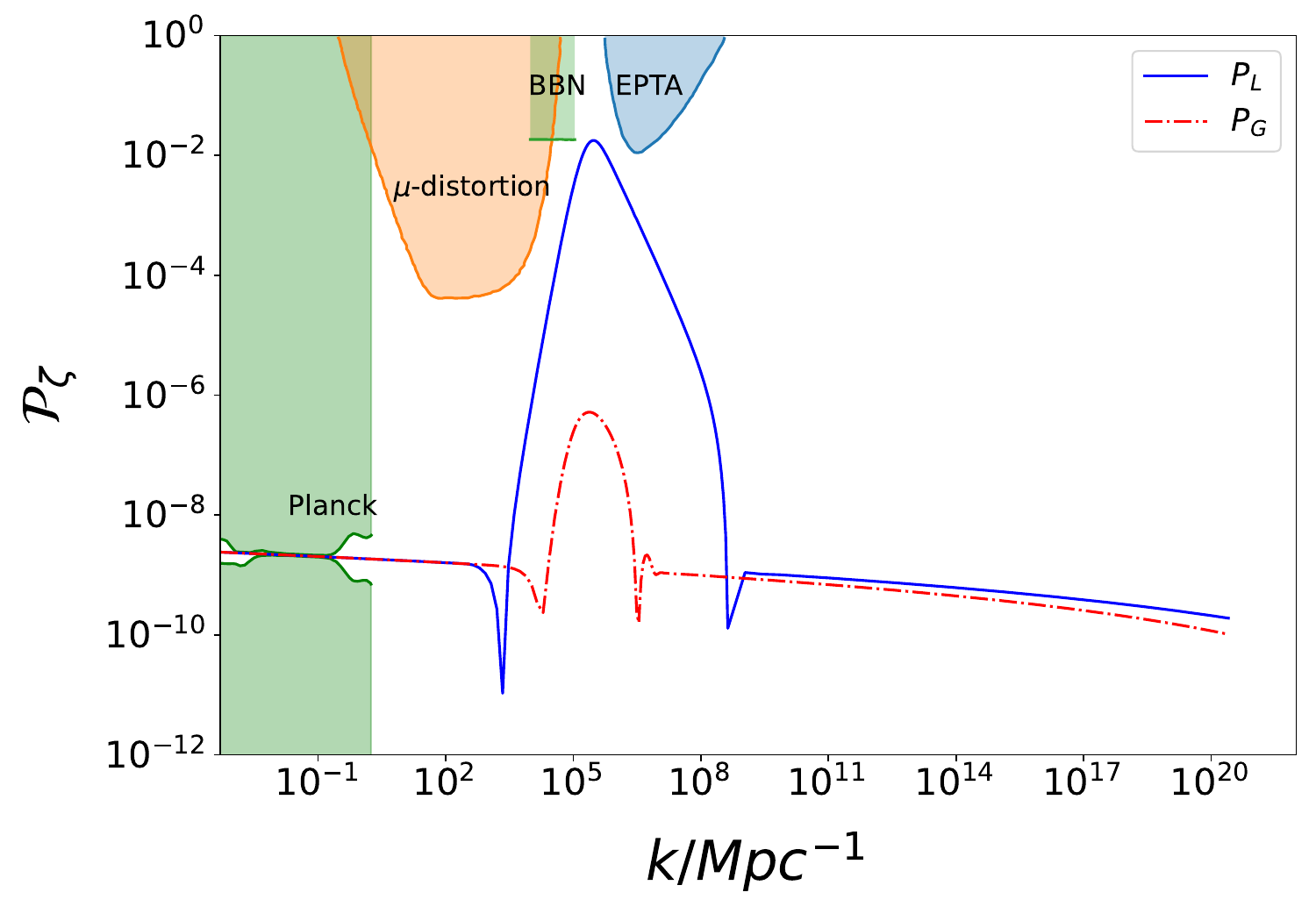}
\caption{The scalar power spectrums under the same parameters $b=4\times10^{-4}, c=0.00999227, \phi_0=5.1$. The $P_L$ and $P_G$ represent the Lorentz and Gaussian cases, respectively. The shaded regions represent the observation constraint \cite{Planck:2018jri, Inomata:2018epa, Inomata:2016uip, Fixsen:1996nj, Chluba:2012we, Jeong:2014gna}. }
\label{Pr-GL1}
\end{figure}

\begin{figure}[H]
\centering
\includegraphics[width=0.45\textwidth]{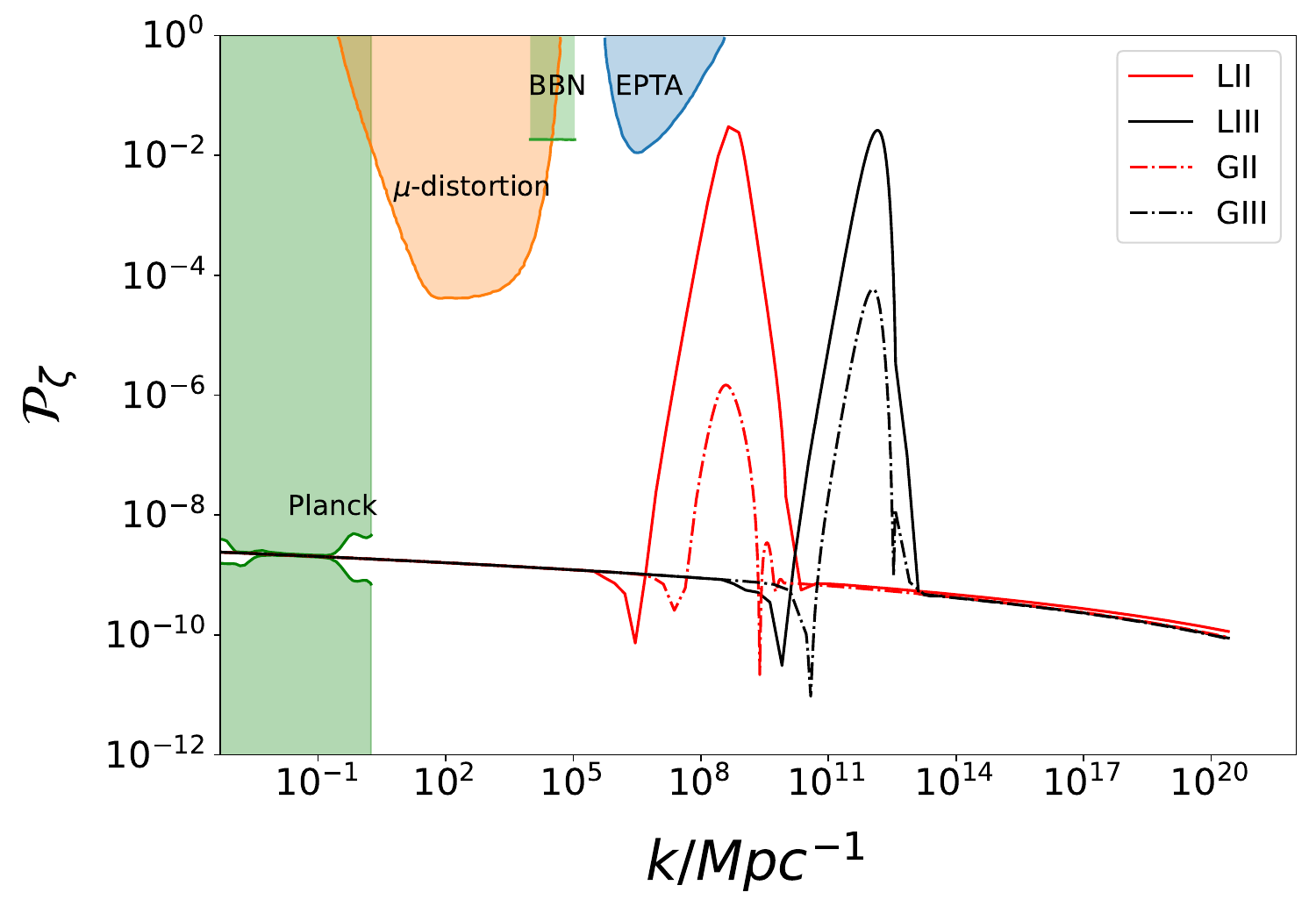}
\caption{The scalar power spectrums for two other parameters (sets II and III). The shaded regions are consistent with Fig. \ref{Pr-GL1}.}
\label{Pr-GL}
\end{figure}

We have clearly illustrated the effects of Lorentz and Gaussian bumps on scalar power spectrum $P_{\zeta}$ in Fig. \ref{Pr-GL1}. For parameter set I $b=4\times10^{-4}, c=0.00999227, \phi_0=5.1$, the peak power spectrum reaches $P_\text{L}\sim\mathcal{O}(10^{-2})$ for the Lorentz bump, significantly higher than $P_\text{G}\sim\mathcal{O}(10^{-7})$ for the Gaussian bump. To verify the generality of this stronger enhancement, we tested two additional parameter sets (II and III in Table \ref{tab:1}). The results consistently show that the Lorentz bump produces a more sufficient enhancement, as shown in Fig. \ref{Pr-GL}.


\begin{table}[H]
\begin{center}
\setlength{\tabcolsep}{1mm}{
\begin{tabular*}{\hsize}{@{}@{\extracolsep{\fill}}cccc}
\hline
$set$ &$b(10^{-4})$ &$c$ &$\phi_{\text{0}}$\\
\hline
$I$  &4 &0.00999227 &5.1  \\
$II$  &4.1 &0.00794544 &4.89   \\
$III$  &4.4 &0.00541278 &4.63  \\
\hline
\end{tabular*}}
\caption{Different sets of parameters in bump function.}
\label{tab:1}
\end{center}
\end{table}

Note that, besides investigating how the form of bump functions in inflation models impacts the primordial power spectrum, in fact, we can also analyze how the parameters b and c are crucial in determining the shape of the power spectrum. As shown in Fig.\ref{Pr-GL1} and \ref{Pr-GL}, the shapes of the power spectrum $P_\zeta$ are similar in both cases. In order to further investigate the physical meaning of b and c, we add an appendix that provides a detailed analysis under inflation constraints and physical settings. Interestingly, the results in the appendix can also offer some clues to distinguish these two bump models.

\begin{table*}[ht]
\begin{center}
\setlength{\tabcolsep}{1mm}{
\begin{tabular*}{\hsize}{@{}@{\extracolsep{\fill}}cccccccc}
\hline
$set$ &$\phi_{\text{e}}$ &$N$ &$k_{\text{peak}}/\text{Mpc}^{-1}$ &$P_{\zeta(peak)}$ &$\frac{M_{\text{peak}}}{M_{\odot}}$ &$Y_{\text{PBH}}^{\text{peak}}$ &$f/\text{Hz}$\\
\hline
$LI$    &0.61435 &67.2 &$3.13\times{10^{5}}$ &0.031 &37.51 &$2.37\times{10^{-5}}$ &$3.66\times{10^{-10}}$ \\
$LII$   &0.61449 &62.7 &$4.45\times{10^{8}}$ &0.03 &$1.186\times{10^{-5}}$ &0.013 &$6.87\times{10^{-7}}$ \\
$LIII$  &0.61391 &61.0 &$1.42\times{10^{12}}$ &0.026 &$1.81\times{10^{-12}}$ &0.72 &$2.19\times{10^{-3}}$ \\
$GI$    &0.61433 &61.9 &$2.49\times{10^{5}}$ &$5.22\times{10^{-7}}$ &59.3 &$0$ &$3.84\times{10^{-10}}$ \\
$GII$   &0.61464 &61.2 &$3.85\times{10^{8}}$ &$1.47\times{10^{-6}}$ &$2.48\times{10^{-5}}$ &0 &$5.95\times{10^{-7}}$ \\
$GIII$  &0.61404 &60.8 &$1.08\times{10^{12}}$ &$5.84\times{10^{-5}}$ &$3.1\times{10^{-12}}$ &0 &$1.67\times{10^{-3}}$ \\
\hline
\end{tabular*}}
\caption{Physical quantities for two bump function cases within different sets of parameters. }
\label{tab:2}
\end{center}
\end{table*}

\section{The PBHs abundance of the Lorentz and Gaussian bumps models}\label{sec:4}

The PBHs can be formed due to curvature perturbations during the inflation process, and their abundance will be enhanced when the scalar perturbation is amplified at a small scale. Note that, the amplification can be clearly illustrated from the power spectrum $P_{\zeta}$. Therefore, there should be some relationship between the abundance and the power spectrum. For simplicity, we here choose the common Press-Schechter (PS) theory~\cite{Press:1973iz}. In this theory, the fractional energy density of PBHs $\beta$ related to the power spectrum is
\begin{align}\label{beta}
\beta&=\frac{\rho_{\text{PBH}}}{\rho_{\text{tot}}}\approx{\sqrt{\frac{2}{\pi}}\frac{\sqrt{P_{\zeta}}}{\mu_{\text{c}}}e^{\left(-\frac{\mu_{c}^2}{2P_{\zeta}}\right)}},
\end{align}
where $\rho_{\text{tot}}$ and $\rho_{\text{PBH}}$ represent the energy density of the cosmic background and PBHs, respectively. The $\mu_{\text{c}}=9\delta_{\text{c}}/2\sqrt{2}$ \cite{Yi:2020cut}, and the threshold of density perturbation for the formation of the PBHs are usually chosen as $\delta_{\text{c}}=0.4$ \cite{Harada:2013epa, Yi:2020cut}.

\begin{figure}[H]
\centering
\includegraphics[width=0.45\textwidth]{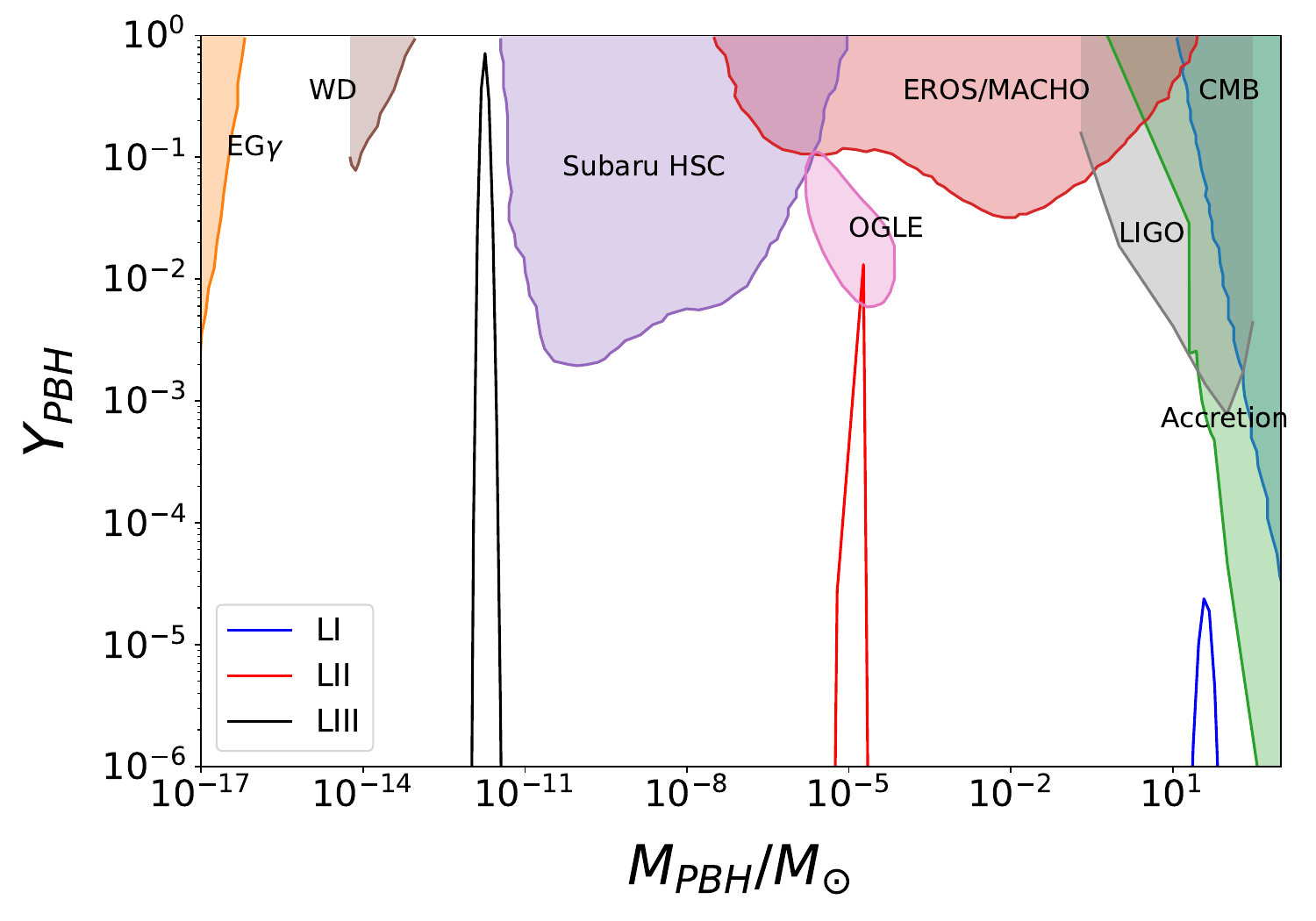}
\caption{The abundances of PBHs as DM produced by the Lorentz bump cases. The dashed line represents the PBH abundance, and the shaded regions represent the constraints on PBHs from various observations\cite{Ali-Haimoud:2016mbv, Poulin:2017bwe, Carr:2009jm, Niikura:2017zjd, Griest:2013esa, EROS-2:2006ryy, Vaskonen:2019jpv, Niikura:2019kqi, Carr:2020gox}.}
\label{Ypbh-M}
\end{figure}

The PBHs abundance $Y_{\text{PBH}}$ related to $\beta$ is given as \cite{Carr:2016drx, Yi:2020cut}
\begin{align}
\label{Y_{PBH}}
Y_{\text{PBH}}\equiv \frac{\Omega_\text{PBH}}{\Omega_\text{DM}} &=\frac{\beta}{3.94\times10^{-9}}\left(\frac{\gamma}{0.2}\right)^{\frac{1}{2}}\left(\frac{g_{*}}{10.75}\right)^{-\frac{1}{4}}\nonumber\\
&\times\left(\frac{0.12}{{\Omega}_{\text{DM}}h^{2}}\right)\left(\frac{M}{M_{\odot}}\right)^{-\frac{1}{2}}.
\end{align}
In this equation, $M$ and $M_\odot$ are the mass of PBHs and solar mass, respectively. The numerical factor $\gamma=0.2$, and the effective degrees of freedom $g_*=107.5$ for $T > 300\,\text{GeV}$ and  $g_*=10.75$ for $0.5\,\text{MeV} < T < 300\,\text{GeV}$ \cite{Yi:2020cut}. In addition, the Planck measurements provide the following value for the DM abundance $\Omega_{\text{DM}}h^2=0.12$ \cite{Planck:2018vyg} with $h=H/100=0.6727 km\cdot s^{-1}\cdot Mpc^{-1}$. The mass M of the PBH that forms during radiation domination is of the same order as the horizon mass $M_H$, $M_{\text{PBH}}=\gamma M_H$. Note that, the mass of PBHs generation is related to the
scalar perturbation k is given by \cite{Yi:2020cut}
\begin{gather}\label{M_{PBH}}
M_{\text{PBH}}=3.68\left(\frac{\gamma}{0.2}\right)\left(\frac{g_*}{10.75}\right)^{-\frac{1}{6}}\left(\frac{k}{10^6 Mpc^{-1}}\right)^{-2}M_\odot.
\end{gather}
After using the above numerical results of the power spectrum, we can calculate the values of $\beta$ and $M_\text{PBH}$ from Eqs.\eqref{beta} and \eqref{M_{PBH}}. Hence, from Eq. \eqref{Y_{PBH}}, we obtain the abundance of PBHs as dark matter in Fig. \ref{Ypbh-M}. Note that, since one set of parameters usually obtains only one peak in the figure, we also plot the abundance of PBHs related to two another sets (II and III) in table \ref{tab:1}. Moreover, for conveniently comparing, we have written some corresponding significant quantities in table \ref{tab:2}.

From table \ref{tab:2}, the abundances of PBHs in the Gaussian bump cases are all nearly zero, and hence there is no observable peak related to the Gaussian bump in Fig. \ref{Ypbh-M}. Clearly, the Lorentz bump case produces a more sufficient abundance of PBHs than the Gaussian bump case. Furthermore, the Lorentz bump case can provide different masses of PBHs with significant observable astronomy. For example, stellar-mass PBHs of approximately $30M_{\odot}$ can be generated for set LI, which might be explained as the LIGO events \cite{Bird:2016dcv, Sasaki:2016jop}. For set LII, our models produce PBHs with the mass $M_{\text{PBH}}\sim 10^{-5}M_{\odot}$, which can explain Optical Gravitational Lensing Experiment (OGLE) ultrashort-timescale microlensing events \cite{Niikura:2019kqi, Mroz:2017mvf} and the anomalous orbits of trans-Neptunian objects \cite{Scholtz:2019csj}.  For set LIII, the PBHs with masses around $10^{-12}M_{\odot}$ can be produced, and its abundance is approximate $Y_{\text{PBH}}\approx1$, and which could make up almost all DM. 

\section{Scalar-induced gravitational waves}\label{sec:5}
It should be pointed out that the SIGWs signal is also generated when the large density perturbations produce PBHs~\cite{ Yi:2020cut, Kawai:2022emp}. Hence, in this section, we will further investigate the different effects of bump functions on the SIGWs signal. The scalar perturbation metric in Newtonian gauge under the cosmological background is written as
\begin{align}
ds^2&=-a^2(\eta)(1+2\mathbf{\Phi})d\eta^2 \nonumber\\
&\quad +a^2(\eta)\left[(1-2\mathbf{\Phi})\delta_{ij}+\frac{1}{2}h_{ij}\right]dx^idx^j, \label{gauge}
\end{align}
where $a(\eta)$ is the scale factor of the universe, $\mathbf{\Phi}$ is the scalar perturbation, and $h_{ij}$ is the transverse traceless tensor perturbation mode. Through the Fourier transformation, the tensor modes can be expressed as  
\begin{align}
h_\text{ij}(\eta,\mathbf{x}) &=\int\frac{d^3k}{(2\pi)^{3/2}}\left[e^{+}_\text{ij}(\mathbf{k})h^+_\mathbf{k}(\eta)+e^{\times}_\text{ij}(\mathbf{k})h^{\times}_\mathbf{k}(\eta)\right] e^{i\mathbf{k}\cdot \mathbf{x}},
\end{align}
where $e^+_\text{ij}$ and $e^{\times}_\text{ij}$ are two basic polarization tensors. After taking the second order perturbations of Einstein's field equations into account, the second order tensor perturbation of $h_k$ satisfies \cite{Kohri:2018awv}
\begin{align}
\label{tensor equation}
h_\mathbf{k}''(\eta)+2\mathcal{H}h_\mathbf{k}'(\eta)+k^2h_\mathbf{k}(\eta)=4S_\mathbf{k}(\eta),
\end{align}
where $^\prime$ denotes $\frac{d}{d\eta}$, and the conformal Hubble parameter is $\mathcal{H}=a'/a$. The source term is written as \cite{Kohri:2018awv}
\begin{align}
S_\mathbf{k}&=\int\frac{d^{3}q}{(2\pi)^{3/2}}  e_{\text{ij}}(\mathbf{k})q_iq_j
\Big[2\mathbf{\Phi}_{\mathbf{q}}\mathbf{\Phi}_{\mathbf{k-q}}+\nonumber\\
&\quad\left(\mathcal{H}^{-1}\mathbf{\Phi}'_{\mathbf{q}}+\mathbf{\Phi}_{\mathbf{q}}\right)
\left(\mathcal{H}^{-1}\mathbf{\Phi}'_{\mathbf{k-q}}+\mathbf{\Phi}_{\mathbf{k-q}}\right)\Big],
\end{align}
where $\mathbf{\Phi_k}$ is the Fourier component of the scalar perturbation $\mathbf{\Phi}$, and it can be calculated from
\begin{align}
\label{phi''}
\Phi_k''+4\mathcal{H}\Phi'_k+\frac{k^2}{3}\Phi_k=0.
\end{align}
The primordial value $\phi_k$ is defined through the relationship with $\mathbf{\Phi_k}$ as
\begin{align}
\mathbf{\Phi}_{\mathbf{k}}=\mathbf{\Psi}(k\eta)\phi_\mathbf{k},
\end{align}
and the transfer function $\mathbf{\Psi}(k\eta)$ for the radiation-dominated period is given by
\begin{align}
\mathbf{\Psi}(k\eta)=\frac{9}{(k\eta)^2}\left(\frac{\sin(k\eta/\sqrt{3})}{k\eta/\sqrt{3}}-\cos(k\eta/\sqrt{3})\right).
\end{align}
Therefore, the power spectrum of the curvature perturbation $P_{\zeta}(k)$ can be determined by the two-point correlation function of $\phi_k$ through
\begin{align}
\langle\phi_\mathbf{k}\phi_{\mathbf{k'}}\rangle=\delta(\mathbf{k}+\mathbf{k'})\frac{2\pi^2}{k^3}\left(\frac{3+3w}{5+3w}\right)^2P_{\zeta}(k).
\end{align}
From this power spectrum, and considering the Eq.\eqref{phi''} into account, we can obtain the energy density of SIGWs in the radiation-dominated period \cite{Kohri:2018awv, Ali:2023moi, Lu:2020diy}
\begin{align}
\label{Omega}
\Omega_{\text{GW}}(k)=&\frac{1}{6}\left(\frac{k}{aH}\right)^2
\int_{0}^{\infty}dv\int_{|1-v|}^{1+v}du\\\nonumber
&\left\{\left[\frac{4v^2-(1-u^2+v^2)^2}{4uv}\right]^2\right.\\\nonumber
&\left.\times I_R^2(u,v,x\rightarrow\infty)\mathcal{P_{\zeta}}(kv)\mathcal{P_{\zeta}}(ku)\right\},
\end{align}
where $u=|\mathbf{k-\widetilde{k}}|$, $v=\widetilde{k}/k$, $x=k\eta$, and $I_R$ is kernel function.

In addition, the frequency $f$ and wave number $k$ of the induced gravitational wave satisfy the following equation \cite{Fu:2019vqc}
\begin{align}
f=1.546\times10^{-15}\frac{k}{1Mpc^{-1}}Hz,
\end{align}
and the present fractional energy density of SIGWs $\Omega_{\text{GW,0}}(k)$ is
\begin{equation}\label{Omega0}
\Omega_{\text{GW,0}}(k)=\frac{\Omega_{\text{r,0}}(k)}{\Omega_{\text{r}}(k)}\Omega_{\text{GW}}(k),
\end{equation}
where $\Omega_{\text{r,0}}(k)=9.17\times10^{-5}$ is the current fractional energy density of radiation, and we choose $\Omega_{\text{r}}(k)=1$ during the period of radiation dominated \cite{Yi:2020cut}. 

\begin{figure}[H]
\centering
\includegraphics[width=0.45\textwidth]{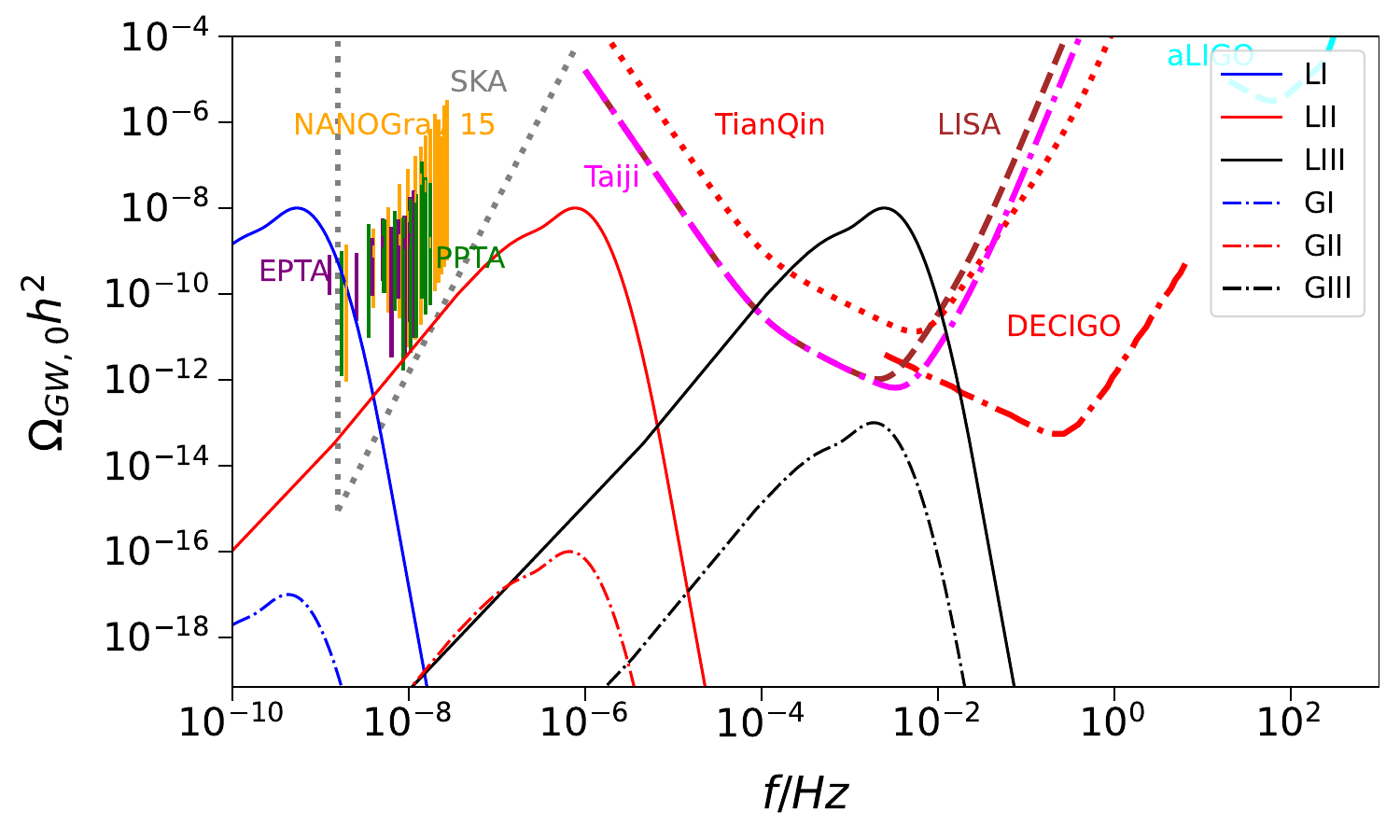}
\caption{The energy density spectrum of SIGWs for the Lorentz and Gaussian cases is plotted by using the three sets in table \ref{tab:1}, which is numerically calculated by the code SIGWfast \cite{Witkowski:2022mtg}. The other curves denote the limits of some GW detectors \cite{LIGOScientific:2016aoc, LIGOScientific:2018mvr, EPTA:2023fyk, Reardon:2023gzh, Moore:2014lga, TianQin:2015yph, Hu:2017mde, Kawamura:2006up, LISA:2017pwj, NANOGrav:2023gor}.}
\label{gw-f}
\end{figure}

In Fig. \ref{gw-f}, we present the curves showing the sensitivity of some GW detectors. For example, the purple dashed, green dot-dashed, and gray dotted curves denote the EPTA DR2full limit \cite{EPTA:2023fyk}, PPTA DR3 limit\cite{Reardon:2023gzh} and the SKA limit\cite{Moore:2014lga}, respectively. The limits of other curves are from the TianQin \cite{TianQin:2015yph}, Taiji \cite{Hu:2017mde}, DECIGO\cite{Kawamura:2006up}, LISA \cite{LISA:2017pwj} and aLIGO \cite{LIGOScientific:2016aoc,LIGOScientific:2018mvr}. In addition, the orange region is the observational result from the North American Nanohertz Observatory for Gravitational Wave (NANOGrav) 15-year data \cite{NANOGrav:2023gor}. We present the energy density $\Omega_{\text{GW,0}}$ of the SIGWs under our three sets of parameters. From this figure, we find that the Lorentz bump case has a higher energy density of SIGWs, and the peak frequency of SIGWs is recorded in table \ref{tab:2}. For example, the gravitational wave signal under a set LI case may be tested by EPTA/PPTA/SKA/NANOGrav. For set LII case may be tested by SKA. For set LIII, the SIGWs may be tested by LISA/Taiji/TianQin/ DECIGO. 

\section{Conclusion and discussion}\label{sec:6}

In this paper, after investigating commonly used Lorentz or Gaussian bump function into base Starobinsky inflation potential function, we obtain different effects of bump function on the formation of PBHs and SIGWs. To clearly illustrate the differences between the two bumps, we have selected the same parameters $b, \phi_0, c$ in bump functions. We find that the Lorentz bump exhibits a stronger ability to enhance the power spectrum at a small scale than the Gaussian bump case, leading to a greater abundance of PBHs as DM. Additionally, the Lorentz bump can provide different masses of PBHs with significant observable astronomy. Furthermore, we also find that the Lorentz bump case shows a higher energy density of SIGWs, making the corresponding gravitational wave signals more detectable in future detectors. 

Note that, the underlying reason of these novel differences may be related to the distinct widths of Lorentz and Gaussian functions in Fig.~\ref{LG}. Therefore, we also further investigate how these parameters in bump functions influence the power spectrum, abundance of PBHs and the energy density of SIGWs in the appendix.  We find that the Lorentz case can generate a sufficiently large abundance of PBHs with stellar masses, while the Gaussian case can not. This difference may distinguish the two cases for practical purposes in future experiments. On the other hand, from making some comparisons by varying the parameters $b$ and $c$ in both two bump functions,  our results confirm the significance of two parameters $b$ and $c$ again, which can change so drastically the behaviour of the power spectrum.

Some issues can be extended in future work. Recent studies reveal that enhancement of small-scale curvature perturbation may cause significant one-loop correction on the CMB scales, which constrains the information mechanism of the PBH \cite{Kristiano:2022maq, Firouzjahi:2023aum}. Therefore, the effect of bump functions on one-loop corrections is an interesting open issue to be further explored. Moreover, the tail of the primordial probability density function (PDF) has been found to decay exponentially rather than Gaussian. Hence, it is attractive to investigate the effect on the PBH abundance from the non-Gaussianity associated with the tails of primordial fluctuations \cite{Kitajima:2021fpq, Pi:2022ysn}. 

In addition, recent studies of peak theory suggest that the statistics of the compaction function provide another method for investigating the formation mechanism of PBHs \cite{Kawai:2021edk, Kawai:2022emp, Germani:2017bcs, Motohashi:2017kbs}. Therefore, both the peak theory and the PS formalism can calculate the mass function of PBH. However, it is worth emphasizing that the PS formalism adopted in our paper is a simplification, one can use the peak theory to discuss the PBH mass function for a more accurate analysis. On the other hand, our assumption of the Gaussianity neglects the potential impact of the non-Gaussianity, which may be crucial for improving PBHs abundance calculations and distinguishing the two bump models. Thus, future work incorporating peak theory and non-Gaussianity would be a valuable extension.

\section*{Acknowledgements}

Ya-Peng Hu thanks Profs. Bum-Hoon Lee, Shi Pi, Ying-li Zhang, and Dr. Zhu Yi for their helpful discussions. We thank the anonymous referees for their valuable comments. This work is supported by the National Natural Science Foundation of China (NSFC) under Grant Nos. 12175105 and 12147175.

\appendix
\section {Comparison results by varying parameters of bump functions}

The paper has shown that under the same parameters, Lorenz has a stronger ability to enhance the power spectrum. Then, if the parameters of both bump functions are not fixed, what influence they will have on the power spectrum might be an interesting topic.
It might give potential insights into distinguishing bump functions.

In the scenario, we fix the initial value of inflation $\phi_{in}$, which corresponds to fixing the inflation initial time. We think this setting may also be reasonable and physical. To be specific, we fix $\phi_{in}=5.42$, which is equivalent to the total e-folding number $N = 60$ for the basic Starobinsky potential without the bump. 

\begin{figure}[H]
\centering
\includegraphics[width=0.45\textwidth]{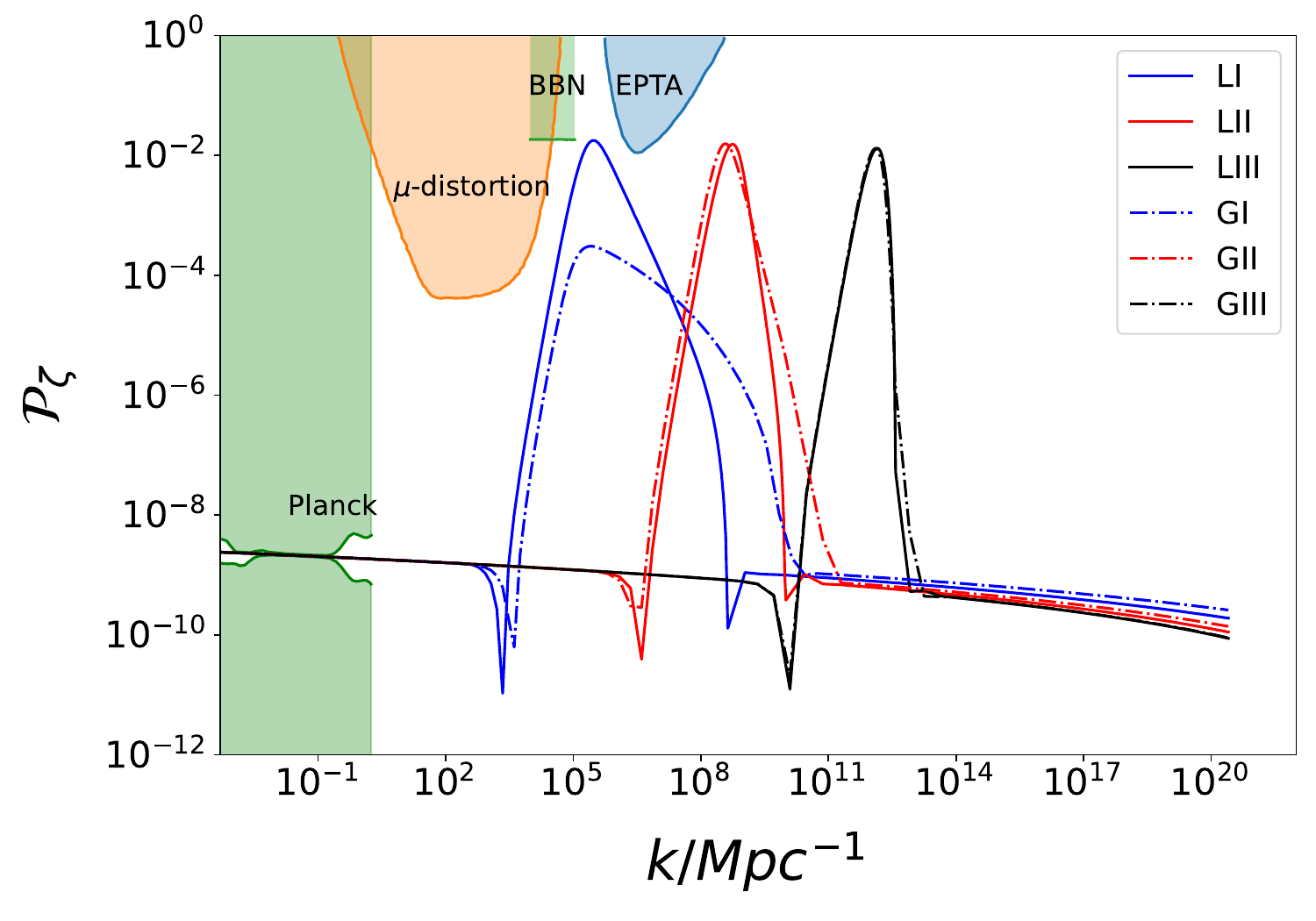}
\caption{ Comparison of the scalar power spectrum for Lorentz and Gaussian cases.}
\label{Pr-A}
\end{figure}

We adjusted the parameters of the two bump functions to enhance the power spectrum. The related quantities are recorded in tables \ref{tab:A} and \ref{tab:B}. The results obtained by Lorenz and Gaussian are consistent with the CMB constraints of large scales in terms of spectral index $n_s\sim0.968$ and tensor-to-scalar ratio $r\sim0.003$. Noteworthy, if the power spectrum of the Gaussian case can reach the order of $O(10^{-2})$ at $k\sim10^{5}$ for the parameters b = 0.000421 and c = 0.0099733 at $\phi_0=5.1$, the total e-folding number before the end of inflation $N\sim79$ does not meet the inflation constraints ($N\leq70$). As demonstrated by GI in Fig. \ref{Pr-A}, under the constraints of inflation, the maximum value of the power spectrum for the Gaussian case at $k\sim10^{5}$ is $O(10^{-4})$. 

\begin{table}[H]
\begin{center}
\setlength{\tabcolsep}{1mm}{
\begin{tabular*}{\hsize}{@{}@{\extracolsep{\fill}}cccccc}
\hline
$set$ &$b(10^{-4})$ &$c$ &$\phi_0$ &$\phi_{\text{end}}$ &N\\
\hline
$LI$  &4 &0.00999311 &5.1 &0.61431 &67  \\
$LII$  &4.1 &0.00770966 &4.89  &0.61403 &62 \\
$LIII$  &4.4 &0.00541278 &4.63  &0.61399 &61 \\
$GI$  &4.21 &0.00999 &5.1 &0.61435 &70 \\
$GII$  &4.13 &0.0077345 &4.89 &0.61466 &64 \\
$GIII$  &4 &0.0044063 &4.63 &0.614394 &61 \\
\hline
\end{tabular*}}
\caption{Model parameters and the numerical results}
\label{tab:A}
\end{center}
\end{table}

\begin{figure}[H]
\centering
\includegraphics[width=0.45\textwidth]{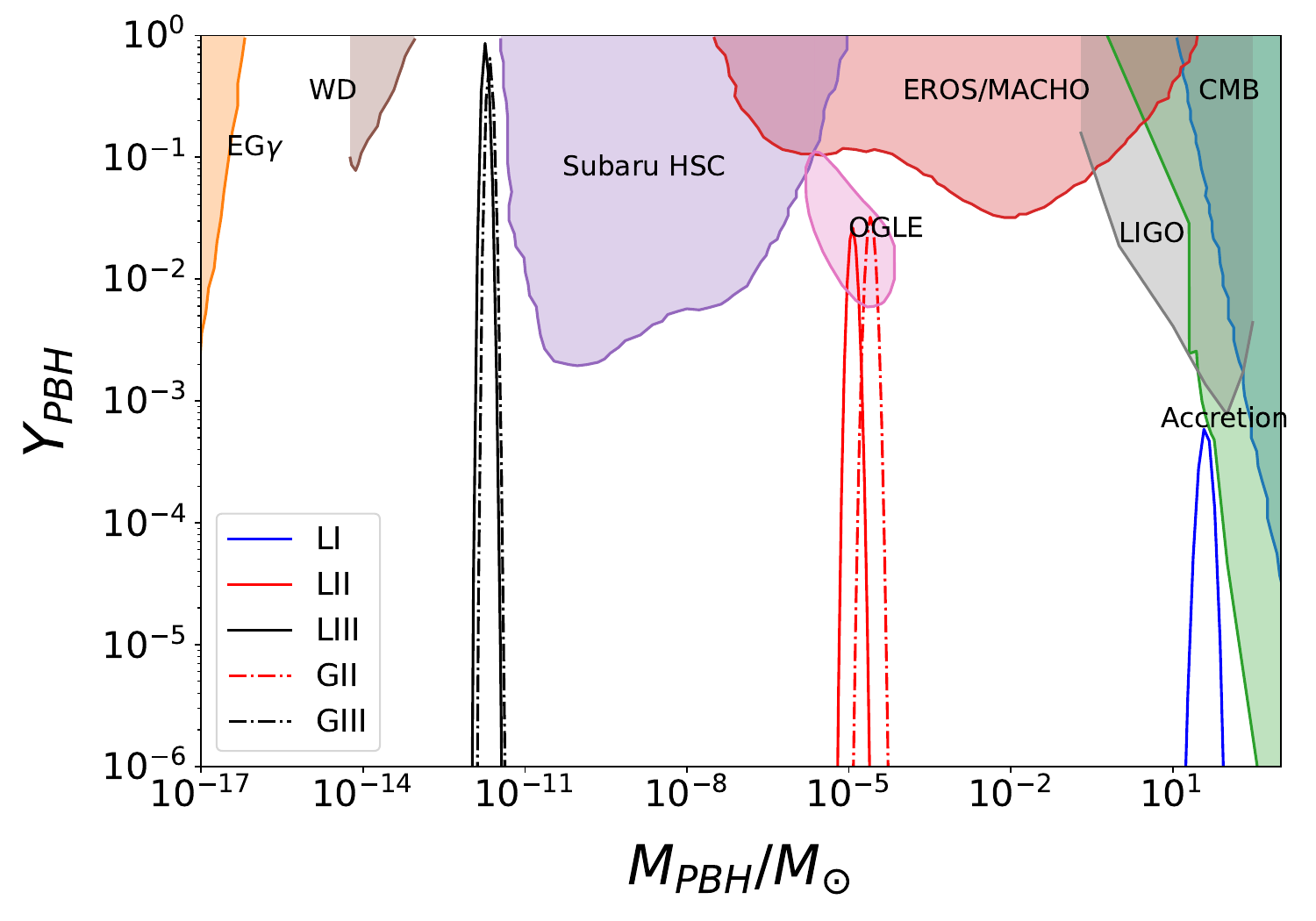}
\caption{The abundances of PBHs as DM  for the Lorentz and Gaussian cases by using the sets in table \ref{tab:A}.}
\label{Ypbh-A}
\end{figure}

\begin{figure}[H]
\centering
\includegraphics[width=0.45\textwidth]{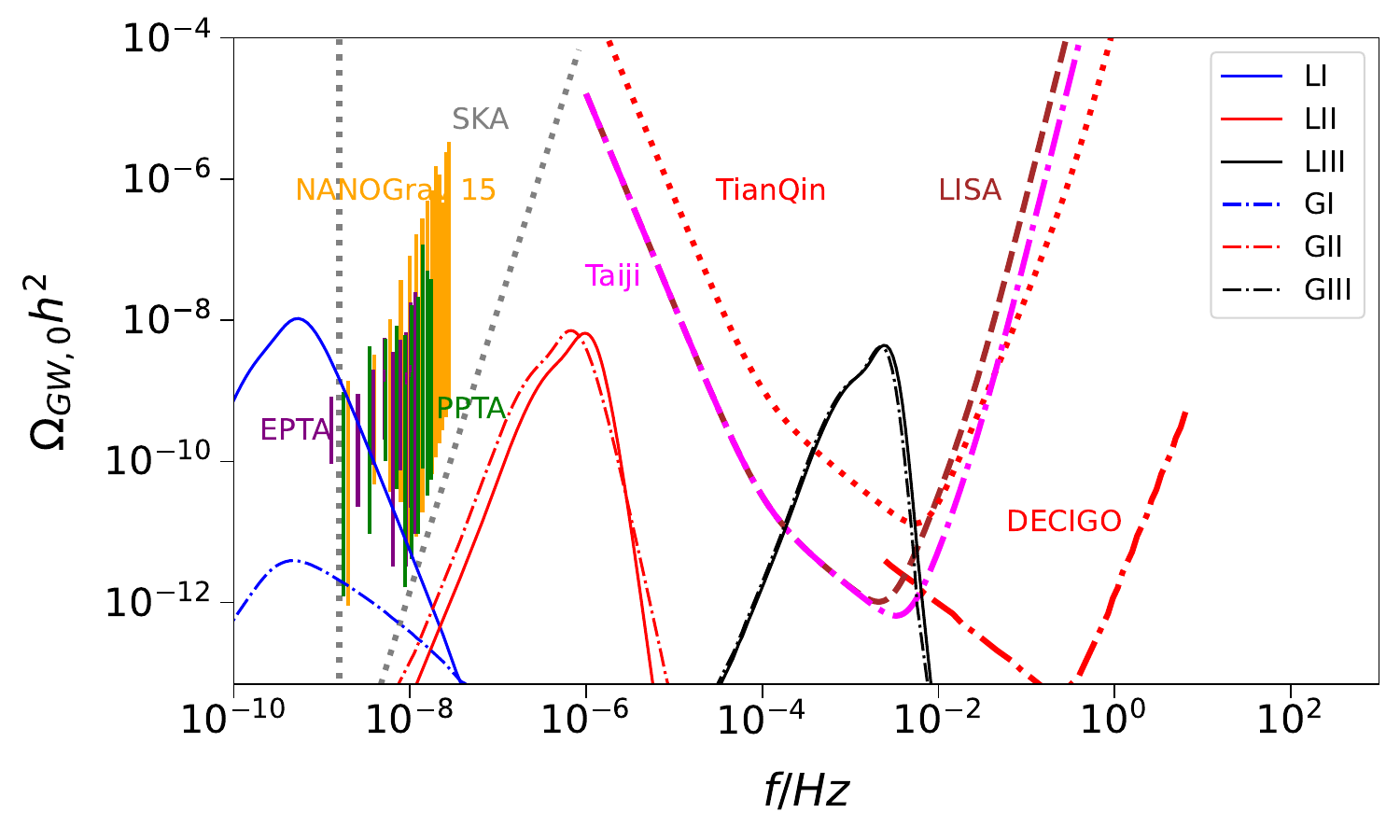}
\caption{The energy density spectrum of SIGWs for the Lorentz and Gaussian cases by using the sets in table \ref{tab:A}}
\label{gw-A}
\end{figure}

To further compare the different effects of the Lorentz bump and the Gaussian bump. In the following, we obtain these abundances of PBHs as DM and the energy density spectrum of SIGWs for the Lorentz and Gaussian cases by using the sets in Table \ref{tab:A}, as shown in Fig.\ref{Ypbh-A} and  Fig.\ref{gw-A}.

Interestingly, from the Fig.\ref{Ypbh-A}, one finds that the shapes of the abundances of PBHs are similar in both cases, which appears to have an apparent degeneracy between the Gaussian and Lorentz bumps related to parameters b and c. However, we also find that the Lorentz case (LI) can generate a sufficiently large abundance of PBHs around the $\mathcal{O}(10^{-4})$ with stellar masses, while the Gaussian case (GI) does not, which gives some clues to remove degeneracy under the local form of Gaussianity cases. Moreover, we obtain a broader frequency range of detectable nHz SIGWs in Fig. \ref{gw-A}. Our results imply that the tiny Lorentz bump in the inflation potential might be advantageous in favor of generating PBH.

\begin{table}[H]
\begin{center}
\setlength{\tabcolsep}{1mm}{
\begin{tabular*}{\hsize}{@{}@{\extracolsep{\fill}}ccccc}
\hline
$k_{\text{peak}}/\text{Mpc}^{-1}$ &$P_{\zeta(peak)}$ &$\frac{M_{\text{peak}}}{M_{\odot}}$ &$Y_{\text{PBH}}^{\text{peak}}$ &$f_{c}/\text{Hz}$\\
\hline
$3.13\times{10^{5}}$ &0.018 &37.51 &$0.00059$ &$4.84\times{10^{-10}}$ \\
$5.57\times10^{8}$ &0.015 &$1.186\times10^{-5}$ &0.026 &$8.61\times{10^{-7}}$ \\
$3.5\times10^{12}$ &0.013 &$1.83\times10^{-12}$ &0.862 &$2.19\times{10^{-3}}$ \\
$2.49\times{10^{5}}$ &0.0003 &59.3 &$0$ &$3.9\times{10^{-10}}$ \\
$3.85\times10^{8}$ &0.016 &$2.48\times10^{-5}$ &0.032 &$6.4\times{10^{-7}}$ \\
$1.28\times10^{12}$ &0.013 &$2.24\times10^{-12}$ &0.65 &$2.2\times{10^{-3}}$ \\
\hline
\end{tabular*}}
\caption{Results for the peak values of scale, the primordial scalar power spectra, peak mass, peak abundance of PBH, and the peak frequency of SIGWs.}
\label{tab:B}
\end{center}
\end{table}

\end{document}